\documentstyle{article}
\input epsf
\catcode`@=11
\def\jimtwoup{\twocolumn\sloppy\flushbottom\parindent 2em
        \parskip .33\baselineskip
        \leftmargini 2em\leftmarginv .5em\leftmarginvi .5em
        \oddsidemargin 0in      \evensidemargin 0in
        \columnsep .4in \footheight 0pt
        \textwidth 10in \topmargin  -.4in
        \headheight 0pt \topskip 0in
        \textheight 6.9in \footskip 30pt
        \hoffset -.5in \voffset -.25in
        \def\@oddfoot{\hfil\thepage\hfil\addtocounter{page}{1}
                \hspace{\columnsep}\hfil\thepage\hfil}
        \let\@evenfoot\@oddfoot \def\@oddhead{} \def\@evenhead{} }
\jimtwoup
\catcode`@=12
\title{Percolation on Strings and the Cover-up of the $c=1$ Disaster}
\author{
Geoffrey Harris\\[1.5em]
Dept. of Physics,\\
Syracuse University,\\
Syracuse, NY 13244, USA\\
\\
  gharris@npac.syr.edu\\[1.0em]}
\begin{document}
\maketitle

\vfill
\begin{abstract}
We study percolation on the worldsheets of string theory for
$c=0,1/2,1$ and $2$.  For $c<1$ we find that critical exponents
measured from simulations agree quite well with the theoretical values.
For $c=1$ we show how log corrections determined from the exact
solution reconcile numerical results with the
KPZ predictions.  We extend this analysis to the large $c$
regime and estimate how finite-size effects will effectively
raise the ground state energy, masking the presence of the tachyon
for moderate values of $c > 1$.   It thus appears likely that simulations
for $c=2,3 \ldots$ on numerically accessible lattices
will fail to even
capture the qualitative behavior of the continuum limit.
\end{abstract}
\vfill
\begin{flushright}
  {\bf SU-HEP-4241-555}\\
  {\bf SCCS 599}\\
  {\bf hep-th/9310137 }\\
\end{flushright}
\vfill
\newpage
\newpage
\section{Introduction}
     Within the last few years, the critical behavior of low-dimensional
($c \leq 1 $) string theories has been understood in considerable detail.
In particular, both analytic and numerical work has yielded
a description of the intrinsic geometry of the worldsheet of these
models \cite{Dist,David,Migdal,Kawai}.
The worldsheet has been shown to be very spiky and branched;
it can be characterized
by a scaling distribution of embryonic universes \cite{JainMat,JainMatAmb}.
As $c$ increases towards $1$, embryonic universes of all
sizes but with necks
of the order of the cutoff proliferate and perhaps saturate the
worldsheet.

    A direct determination of properties of
the intrinsic geometry of string theories
constructed from $c > 1$ matter has proven much more difficult.
There are many theoretical indications that the worldsheet
degenerates into polymer-like configurations above $c = 1$.  Though
the $c < 1$ surfaces are quite spiky, they still appear to
retain some of the character of two dimensional surfaces;
this perhaps is no longer true for $c > 1$.  For
at $c = 1$ the dressed identity operator (coupled to the cosmological
constant) becomes tachyonic.  Above $c=1$, this
operator should then spawn states which are non-local in the
worldsheet metric that
effectively tear the worldsheet apart \cite{Sei,GinMoo}.
Related computations in
Liouville theory indicate that a sort of Kosterlitz Thouless transition
is anticipated at $c = 1$, in which vortex configurations of the Liouville
field become unbound \cite{Cates}.  Unfortunately though, this picture
has not been verified via an exact calculation for $c > 1$;
the appropriate matrix
models have not been amenable to exact solution (though see
\cite{HikBre} and \cite{Wexler}) and Liouville theory yields
complex exponents in this regime.  We would therefore like to
detect, through Monte Carlo simulations, some characteristic
of the geometry of worldsheets that distinguishes $c \leq 1$
from $c > 1$ and signals the onset of the tachyon.

   In fact, numerical work has not provided any evidence
of a dramatic change in the internal geometry of the worldsheet
as $c$ passes $1$.  Simulations, albeit on
modest-sized lattices,  only have shown convincing evidence of
branched-polymer structure on the worldsheet for $c > 10-12$.
Recently, there has been much work on simulations of multiple
Potts models coupled to gravity \cite{Ambjmult,Catmult,Bailmult};
the systems studied generally have small values of $c$ both
greater and less than $1$.   In these studies, the internal geometry
was characterized by measuring the distribution of
discrete curvature, or more precisely, the
distribution of ring lengths on $\phi^3$ graphs dual to triangulations.
These measurements indicated that no dramatic change in
these curvature distributions occurs as c is increased above $1$.
Interpreting these results rigorously is difficult, though, since
it is not clear how to identify these curvature distributions with
continuum (universal) correlation functions.

    To search for a transition to branched polymers, Ambj{\o}rn et al.
\cite{Ambjmult} also simulated Ising models on typical fixed
random triangulations sampled from the set of worldsheets characteristic
of gravity coupled to matter of central charge $c$.  Each
triangulation was generated during a simulation of a multiple Potts
or Gaussian models coupled to gravity; the Ising model was then
simulated on each of these particular triangulations.
These authors showed that the Ising model on a branched polymer
will not undergo a phase transition.  Only when $c$ reached
or exceeded $12$ did they find that no finite-temperature
Ising transition was evident on the surfaces extracted from their
simulations.

  For fairly low central charge,
the coupling between matter and gravity, as inferred from the
above numerical work,
{\it{appears}} to be fairly weak, perhaps too weak to
drive the worldsheet into a branched polymer phase \cite{Ambjmult}.
It has thus been suggested
that worldsheets of gravity coupled to matter with
$c > 1$ but less than about $10$ might lie in some sort of intermediate
phase and not undergo a transition to branched polymers for $c$ above
$10$ or so \cite{Davidrev,Ambjmult}.
  Such a scenario is quite intriguing,
especially since the existence of such a phase does not
seem to be anticipated by Liouville theory.

   To reveal the character of the critical geometry of string
worldsheets, we shall examine site percolation
on dynamical triangulations coupled to matter.
The distribution of percolation clusters
should serve as a sensitive probe of the branched structure
of the worldsheet, distinguishing $c \leq 1$ from a branched
polymer phase.  Essentially, embryonic universes act as traps
that prevent percolation clusters from growing, while regions of
high connectivity will enhance the span of percolating clusters.
We also indicate how the critical behavior of percolating
clusters on worldsheets
of $c<1$ matter coupled to gravity can be determined from Liouville
theory.  The quality of the results of numerical simulations of
two-dimensional gravity has often been called into question,
with the suspicion
that the lattices used were often too small too even qualitatively
reflect continuum behavior.  We shall see however that we can reproduce
on lattices of several thousand nodes,
with quite good precision (of order $1\%$), the theoretical predictions
of these percolation exponents for $c < 1$.
At $c=1$, matrix model solutions and Liouville theory
\cite{Klebanov} predict a
more complicated logarithmic dependence of correlation functions
on the cosmological constant.  We also find that our results
are consistent with the functional dependence predicted by matrix
models along with the exponents of Liouville theory.
Our data shows only an apparent weak coupling of $c=1$ matter to
gravity, but this is misleading.  The logarithmic
scaling of the matrix model
implies that effectively $c=1$ matter only has a small effect
on the percolative structure of the lattice for small lattice
sizes.  For larger lattices however, the matrix model solutions predict
that  the influence of the matter
coupling to gravity should gradually become more pronounced.

        We also will demonstrate that percolation on
non-interacting
branched polymers does not undergo a transition for $p < 1$.
We shall then find evidence suggesting that for $c=2$, the critical
behavior of percolating clusters (or indeed perhaps the lack of criticality)
may be qualitatively different from that of $c <  1$.  The $c=2$
observations,
which need to be interpreted with great care,
{\it{at first glance}} appear to be
indicative of a breakdown of the behavior that characterizes percolation
on surfaces.

    In illuminating the above results, we shall argue that for $c$ somewhat
greater than one, the finite size of the lattice will largely mask
the theoretically anticipated degeneration of the worldsheet.
The infrared cutoff should induce a large shift
in the tachyonic energy, preventing the unrestrained
proliferation of tachyons
on not too large lattices.  We can, in fact, estimate the magnitude
of this effect.  With this in mind, we then argue that
for $c$ somewhat greater than $1$, we are most likely
measuring finite-size artifacts; yet
the presence and qualitative nature of these effects can be understood
in the context of string theory.

     A delicate analysis of percolation on dynamically triangulated
lattices is possible largely due to a curious but elementary
graphical property of
triangulations: they are self-matching.  This constrains the
value of the percolation threshold to be $1/2$, with a
few caveats.  We shall explain this further in Section 3.

    In a companion paper \cite{ournewpaper}, an analysis
of the critical properties of the two-species Ising model coupled
to two-dimensional gravity appears.   As in this paper, it is
essential to incorporate logarithmic corrections to scaling to
compare numerical results with theory.  Percolation is also
simulated on random lattices coupled to two Ising species, yielding
results consistent with those presented here.

     The plan of this paper is as follows.  We begin by
reviewing basic facts about percolation and discuss how the
KPZ \cite{KPZ} equation predicts the behavior of percolating
clusters on $c < 1$ dynamical lattices.  We then explain the theoretical
predictions for $c=1$ and also analyze the behavior of
percolation on branched polymers.  A discussion of the constraint
$p_c = 1/2$ follows.  We then describe the numerical
techniques we have used, including an analysis of auto-correlations.
We next present the results of our simulations (for $c = 0, 1/2,
1$ and $2$) and then give our conclusions.

\section{Theoretical Predictions}

      We shall present the results of site percolation on random
lattices sampled from simulations of quantum gravity coupled to matter.
Site percolation describes the following process:  we randomly color
sites `black' on a lattice with probability $p$.  Adjacent
black sites are then connected together to form clusters.
We shall generally be interested in features of the distribution
$n(s)$ of these
clusters as a function of the number of constituent sites $s$.
Typically, we will measure the mean cluster
size, denoted by $\cal{S}$. If one randomly chooses a black point on the
lattice, then on average it will belong to a
cluster of size $\cal{S}$.  Explicitly,
${\cal{S}} = \langle s^2 \rangle / \langle s \rangle$, in which
averages are taken over the distribution $n(s)$.
If a percolation transition occurs at some value $p = p_c$, then
in the infinite volume limit, $\cal{S}$ remains finite for $p < p_c$ but
diverges (it is proportional to the number of lattice sites $N$) for
$p > p_c$ due to the presence of an infinite cluster.
The mean size can also be written as the integral of the pair-connectedness
function
\begin{equation}
    {\cal{S}}_N = {\frac{1}{N}}\sum_{i,j=1}^{N} \langle \delta_{{\cal{C}}_i,
               {\cal{C}}_j} \rangle;
\protect\label{paircon}
\end{equation}
the quantity within the bra-ket is $1$ when $i$ and $j$ lie in
the same cluster and $0$ otherwise.

      One can also consider bond percolation, in which bonds
of the lattice are randomly colored black with probability $p$ and
clusters are then constructed from sites connected by
black bonds.  Bond percolation is generically in the same universality
class as site percolation.  A bond percolation problem on a lattice
$L$, for instance, can be mapped exactly to site percolation
on the covering lattice of $L$ (as defined in reference (\cite{FishEss})).
The properties of bond percolation clusters can be derived
by appropriate averaging with the partition function
\begin{equation}
Z = \sum_{colorings}p^{b}(1-p)^{N_b - b}q^{N_c}
\protect\label{qPottspart}
\end{equation}
with $q \rightarrow 1$; $b$ equals the number of black bonds that constitute
$N_c$ clusters in a lattice
with $N_b$ total bonds.
In fact, the partition function of the $q$-state Potts model on this lattice
can be recast as (\ref{qPottspart}); thus percolation is a
non-interacting limit of these theories.
The $q \rightarrow 1$ Potts model coupled to gravity has been
studied analytically as a matrix model by Kazakov {\cite{Kazakov}}.
This mapping provides
an interpretation of typical percolative properties in the
language of spin models.  For instance,
the relation (\ref{paircon}) is
analogous to the equivalence of the susceptibility ($\cal{S}$)
and the integral of the spin-spin correlation function.
The scaling behavior of this correlation function can then be
computed exactly in $2$ dimensions via a mapping of the Potts model
to the Coulomb gas \cite{DeNijs}.  The result is
\begin{equation}
         \langle \delta_{{\cal{C}}_i,{\cal{C}}_j} \rangle \sim
         |{\vec{r}}_i- {\vec{r}}_j|^{-2(\Delta_{\sigma,q=1}^{o}
               + {\bar{\Delta}}_{\sigma,q=1}^{o})},
\end{equation}
with $\Delta_{\sigma,q=1}^{o} = {\bar{\Delta}}_{\sigma,q=1}^{o}  = 5/96$
\footnote{The weight $\Delta_{\sigma,q=1}^{o}$ is {\it{not}}
identified with a local operator in percolation theory; it just
determines the scaling behavior of a certain class of correlation
functions.}.

    The random surfaces which we shall examine are triangulations
of two-dimensional geometries.  The discretization we use consists
of a sum over all possible triangulations of $N$ vertices, excluding
degenerate triangles \footnote{i.e. those with loops of length $1$ or
$2$ or vertices with fewer than $3$ neighbors},
with weights determined by the partition
function
\begin{equation}
     Z_N = \sum_{T\in{\cal{T}}_N} \rho (T)\exp ( -{\cal{H}}_{matter} );
\protect\label{partfn}
\end{equation}
${\cal{H}}_{matter}$ is the Hamiltonian for Ising or Gaussian
fields and $\rho (T)$ is the measure in the space of triangulations
${\cal{T}}_N$ of $N$ vertices.
 Most of the relevant theoretical calculations are performed
in the grand-canonical ensemble, with the partition function
\begin{equation}
     Z(\mu) = \sum_{N=1}^{\infty} Z_N \exp (-\mu N)
\end{equation}
dependent on $\mu$, the cosmological constant.
The integrated pair-connectedness correlation function
in the grand-canonical ensemble
then satisifies
\begin{equation}
\protect\label{suscgcan}
      \langle \sum_{i,j} \delta_{{\cal{C}}_i,{\cal{C}}_j}
\rangle (\mu) Z(\mu) = \sum_{N=1}^{\infty} N{\cal{S}}_NZ_N \exp (-\mu N).
\end{equation}

    We will thus
simulate the tensor product of the $q=1$ Potts model
with the matter theory coupled to gravity.  Only, since
percolation is not dynamical, it does not generate any back-reaction
on the gravity or the matter.  Using standard arguments
\footnote{applied to the integrated two point function rather
than the one point function as usually presented},
(\cite{DDK}) one can express the scaling behavior of the integrated pair-
connectedness correlation function
of percolation coupled to gravity via the KPZ formula (\cite{KPZ}).
The weight $\Delta_{\sigma.q=1}^{o} = {\bar{\Delta}}_{\sigma,q=1}^{o}
 = 5/96$
is dressed by gravity in a theory
of central charge $ c = c_{matter} \leq 1$ ($c_{percolation} = 0$)
so that
\begin{equation}
\protect\label{KPZmsz}
\Delta_{\sigma,q=1} - {\frac{5}{96}} = \left(1 +
{\frac{1}{12}}\sqrt{1-c}(\sqrt{1-c} - \sqrt{25-c})
\right)\Delta_{\sigma,q=1}
(1 - \Delta_{\sigma,q=1}).
\end{equation}
      This dressed weight determines the scaling of
the integrated pair-connectnedness function with $\mu$
on surfaces of genus $h$:
\begin{equation}
\protect\label{suscmusca}
      \langle \sum_{i,j} \delta_{{\cal{C}}_i,{\cal{C}}_j}
\rangle  (\mu)  Z(\mu) \sim (\mu - \mu_c)^{2(-1 + \Delta_{\sigma,q=1}
) + (2-{\gamma}_s)(1-h)}
\end{equation}
(the $-1$ before the dressed weight
accounts for the integrations of $i$ and $j$
over the surface)
with
\begin{equation}
     \gamma_s = {\frac{1}{12}}\left( c - 1 - {\sqrt{(25-c)(1-c)}}\right).
\end{equation}
Then the relation (\ref{suscgcan}) and
\begin{equation}
Z_N \sim N^{-1 + (\gamma_s - 2)(1-h)}
\end{equation}
yield the finite-size scaling relation
\begin{equation}
\protect\label{mszfss}
  {\cal{S}}_N \sim N^{1 - 2\Delta_{\sigma,q=1}};
\end{equation}
this is the scaling law that we shall verify numerically.
As more generally in spin models, the above exponent is
referred to as $\gamma/\nu d_{H}$.  The mean cluster
size satisfies ${\cal{S}} \sim (p-p_c)^{-\gamma}$, the
correlation length (governed by the decay of the
pair-connectedness function) $\xi \sim (p-p_c)^{-\nu}$
and $d_H$ is the intrinsic Hausdorff dimension of the random surface
being considered.
  We also shall measure the fractal dimension of the largest
cluster at $p_c$; the average maximal size cluster of each
configuration ${\cal{M}}$ scales
as
\begin{equation}
{\cal{M}} \sim  N^{\frac{d_f}{d_H}}.
\end{equation}
 Standard scaling arguments \cite{Staufferbook}
relate $\gamma/ \nu d_H = 2d_f/d_H - 1$.

\subsection{$c=1$}

   The scaling relations become more complicated for $c=1$.
The analytic solutions of the $c=1$ matrix models \cite{c=1solvers} and
a careful analysis of Liouville theory show that correlation
functions no longer scale simply as powers of the cosmological
constant $\mu$.  Instead, the appropriate scaling variable
is $\eta$ which satisfies
{\footnote{
Without loss of generality, we set $\mu_c = 0$.}}
\begin{equation}
\protect\label{logscaling}
\mu = -\eta \ln (\eta) + c_1 \eta + \cdots
\end{equation}
in the limit
of small $\eta$; $c_1$ is a constant that we do not
specify.
The analytic solution
employs a modified Gaussian propagator on phi-cubed lattices
and is thus not based on
the same discretization that we use in our simulations.
We shall assume that the asymptotic scaling relation also holds
for the model we simulate; we will not presume universality
for the subleading coefficient $c_1$.  We therefore
conjecture that the scaling relation (\ref{suscmusca})
should be modified so that
\begin{equation}
\protect\label{suscetasca}
      \langle \sum_{i,j} \delta_{{\cal{C}}_i,{\cal{C}}_j}
\rangle (\mu) Z(\mu) \sim \eta(\mu)^{2(-1 + \Delta_{\sigma,q=1}
) + (2-{\gamma}_s)(1-h)}.
\end{equation}
    Our simulations will be done on worldsheets of toroidal topology
($h=1$)
for which $Z(\mu) \sim \ln (\eta)$ for $c=1$.  To extract
the asymptotic scaling behavior of ${\cal{S}}_N$, we invert the relation
between $\eta$ and $\mu$ order by order in $1/\ln \mu$
and $\ln (-\ln (\mu))/\ln (\mu)$ to obtain
\begin{equation}
\protect\label{inversion}
\eta = -{\frac{\mu}{\ln \mu}}\left( 1 + {\frac{\ln (-\ln \mu)}{\ln \mu}}
+ \left({\frac{\ln (-\ln \mu)}{\ln \mu}}\right)^2
- {\frac{\ln (-\ln \mu)}{(\ln \mu)^2}}
+ \cdots \right) .
\end{equation}
We then expand the inverse Laplace transform of (\ref{suscetasca})
to
obtain
\begin{eqnarray}
\protect\label{susczn}
N{\cal{S}}_NZ_N \sim {\frac{1}{N(N\ln N)^{\omega}}}\left(   1 -
{\frac{\ln \ln N}{\ln N}} + (\frac{\ln \ln N}{\ln N})^2 -
   \frac{\ln \ln N}{
(\ln N)^2} + \cdots \right)^{\omega} \times \nonumber \\  \left( 1 +
  {\frac{\omega \Psi(-\omega)}{\ln N}}
 - \omega\Psi(-\omega){\frac{\ln \ln N}{(\ln N)^2}} + \cdots \right);
\end{eqnarray}
$\omega = 2(-1 + \Delta_{\sigma,q=1}) = -\gamma/\nu d_H - 1$ and $\Psi$
is the digamma
function.
The scaling behavior of $NZ_N$ is obtained by inverse
Laplace transforming $\partial Z(\eta(\mu))/\partial \mu$:
\begin{equation}
\protect\label{partc1}
Z_N \sim {\frac{1}{N}}(1 + {\frac{1}{\ln N}} - {\frac{\ln \ln N}{(\ln N)^2}}
 + \cdots ).
\end{equation}
In addition to the higher order terms that we have dropped from
the inversion, there are additional corrections to the above
formulae.  The corrections to the logarithmic renormalization of
$\mu$ which depend on $c_1$ in (\ref{logscaling})
should lead to contributions to (\ref{susczn}) and (\ref{partc1})
that are competitive with
the smallest corrections we have shown.  We neglect the
usual corrections to scaling that are suppressed by
(non-integer) powers of $1/N$.
To compare this finite-size
scaling form with our data, we shall present a plot of
$\ln ({\cal{S}}_{2N}/{\cal{S}}_N)/\ln 2$ computed from  (\ref{susczn}) and
(\ref{partc1}).   The leading logarithmic contributions
that we have derived above will turn out to be very significant \footnote{
The essential role of
logarithmic corrections in interpreting
numerical measurements of $\gamma_s$ at $c=1$ has been
previously discussed in references \cite{Ambgs,JainMat}.}.

\subsection{$c \gg 1$}

   We are particularly interested in using percolation to uncover
the character of the internal geometry of surfaces for $c > 1$.
In this regard, we discuss the behavior that we would anticipate
in the large $c$ limit.  Ambj{\o}rn, Durhuus, Fr\"{o}hlich and Orland
\cite{Ambdfo}
have argued that in the large $c$ ($d$) limit, the string path integral
(with appropriate measure)
should be dominated by surfaces that form
a branched-polymer like structure.   To show
this, they essentially perform a saddle-point evaluation
of the string path integral
\begin{equation}
\protect\label{gausspartfn}
   Z = \sum_{T \in {\tilde{\cal{T}}}_n} (\prod_i q_i)^{d/2 + \alpha} \int
    \prod_{i,\mu}dX_i^{\mu} \exp \left(-
    \sum_{i,j} C_{ij}\sum_{\mu=1}^{d} (X_i^{\mu} - X_j^{\mu})^2
    \right),
\end{equation}
with the integration over the center of mass of ${\vec{X}}$ implicitly
omitted.  ${\tilde{\cal{T}}}_n$ represents the space
of triangulations that do not
contain loops of unit length or vertices of order less than $3$.
As usual, $C_{ij}$ is the adjacency matrix of the triangulation $T$.
For $i \ne j$, it equals the number of links connecting $i$ and $j$;
$C_{ii} = -q_i$; $q_i$ is the coordination number of site
$i$, equal to the number of links emanating from $i$.
   We can integrate out the Gaussian fields, obtaining
$(\det 'C_{ij})^{-d/2}$ (with the zero-mode omitted).  It can
be shown \cite{Ambdfo} that this
determinant
is minimized
by surfaces consisting of tetrahedra glued together along a common
pair of vertices.
\footnote{This does not rule out the possibility of other saddle point
solutions, though.}.
To perform the gluing, the link connecting these vertices within
each tetrahedron is slit into
two edges; each edge is then glued to a slitted edge of the adjacent
tetrahedron.  This process creates loops of length $2$ which are allowed
in ${\tilde{\cal{T}}}_n$, these are
excluded from the set of triangulations ${\cal{T}}_n$ that we consider
in our simulations.  It has been shown that in $d=0$ and $-2$,
the critical properties are independent of whether loops of length $2$
are included.  Let us assume, as in reference ({\cite{Ambdfo}) that
this is also true in the large $d$ limit, since the configurations
that dominate the saddle point in the space ${\cal{T}}_n$ are quite
complicated and not amenable to the following simple analysis.
We shall then analyze a class of triangulations discussed by
Ambj{\o}rn et al. {\cite{Ambjmult}}
that includes these saddle point solutions; this
class is formed by linearly joining the linked tetrahedra that minimize
the determinant.  A typical triangulation
is depicted in figure~\ref{tetra}.
\begin{figure}
\epsfxsize=4.5in \epsfbox{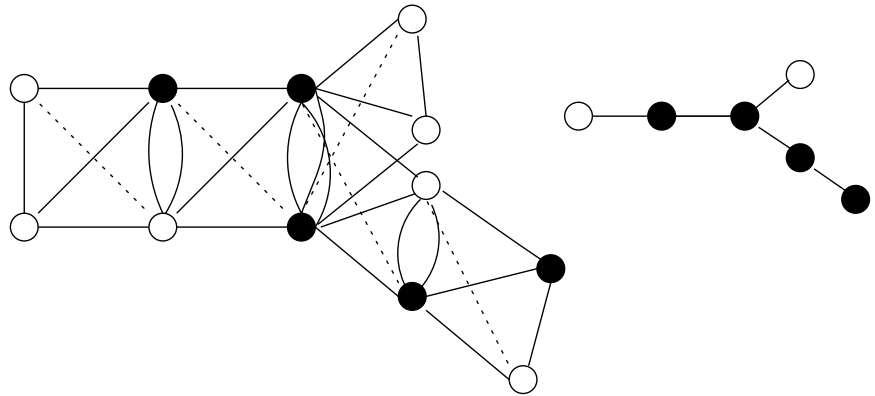}
\protect\caption{\protect\label{tetra}
A sample triangulated surface built from
linked tetrahedra, with vertices colored as in percolation.
The arcs represent slitted edges.  To the right, the associated
branched polymer appears; each tetrahedron corresponds to a polymer
link.}
\end{figure}

         We first show that the percolation problem on this
class of triangulations can be mapped to a
percolation problem on the model of branched polymers
considered by Ambj{\o}rn et al..
Each tetrahedron will correspond to a bond in a branched polymer.
We identify an internal polymer vertex with a pair of sites
connected by a slitted edge on the original
triangulation.  Vertices on the boundary of the polymer are associated
with pairs of sites that lie adjacent at dead-ends of the triangulation.
Now, if either
of the two sites along a slit edge are colored black, then the
percolation cluster to which they belong must include any of the black
sites in adjacent tetrahedra.  If an infinite percolation cluster
is to wind through the lattice, its extent depends only on
whether or not at least one of
two sites on the original triangulation,
associated with a polymer
vertex, is black.  Therefore, we identify black-black, black-white
and white-black pairs of sites with a black polymer vertex;
a white-white pair is mapped to a white polymer vertex.  If
we color sites on the triangulation
with probability $p^*$, then at least one of a pair of sites
is black with probability $ p(p^*) = 1 - (1 - p^*)^2$.  Thus
if the percolation threshold on the original lattice
is $p^*$, then the associated branched polymer will undergo a transition
at $p(p^*)$.

     We now show that these branched polymers (in their critical limit)
do not admit a percolation transition for $p < 1$.
They are described by the partition function
\cite{Ambdfo}
\begin{equation}
Z = \sum_{N} \sum_{t \in t_N} ({\prod}_i w(q_i)) \exp (-\mu N ),
\end{equation}
in which $t_N$ represents the set of trees (graphs with no loops)
of $N$ vertices and each vertex with coordination number $q_i$
is given a weight $w(q_i) = \prod_i(q_i)^{\alpha}$ (since
the determinant
of the adjacency matrix on these surfaces just effectively shifts
the exponent in the measure).

     The analysis of percolation on these polymers proceeds
as in the case of the Bethe lattice \cite{Staufferbook}.  First,
we randomly select a site $i$ on the lattice and a branch
that emanates from it.  Let $T$ denote the mean number of sites
in this branch that belong to the same cluster as $i$.  If the
neighbor $j$ to site $i$ in this branch has coordination number
$q_j$ then
\begin{equation}
    T =  p + p\langle q_j - 1 \rangle T;
\end{equation}
the first contribution $p$ is the probability that site $i$ is black
and in a cluster; if site $j$ is black (with probability $p$) then
each of its $q_j - 1$ subbranches contributes on average $T$ sites.
Therefore,
$T = p/(1 - \langle q - 1\rangle p)$.
If we then randomly select a black site on the branched polymer,
we see that on average it is part of a cluster of size
\begin{equation}
\protect\label{meansizbp}
{\cal{S}} = 1 + \langle q \rangle T = 1 + {\frac{\langle q
\rangle p}{(1 - \langle q-1
\rangle p)}},
\end{equation}
 where ${\cal{S}}$ is the
mean cluster size.    Hence percolation will occur on these trees
as long as the density of dead ends ($q=1$ vertices) is not large
enough to reduce $\langle q \rangle$ below $2$ and
trap all percolating clusters.  If $\langle q \rangle > 2$,
then the mean cluster
size behaves as if we were just considering a Bethe lattice.

     Now we shall compute the mean coordination number
in the critical limit for this model of non-interacting
branched polymers.  First, note that
a polymer vertex of coordination number $q$ is associated with
two sites of the original triangulated surface with coordination
number $3q$.  If this surface is of infinite size and of finite genus,
then simply $\langle 3q \rangle  =6 $.  One can also show in a slightly
more
painstaking fashion that this condition holds in the grand-canonical
ensemble by analyzing
the solution of the Schwinger-Dyson equations for
$Z$ analyzed by Ambj{\o}rn et al. {\cite{Ambjmult}}.  The equations
are written down for `rooted' branch polymers, which emanate from
a marked point with coordination number $1$, they are represented
graphically in figure~\ref{schwing}.
\begin{figure}
\epsfxsize=4.5in \epsfbox{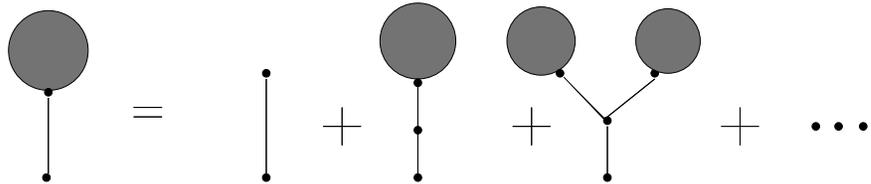}
\protect\caption{\protect\label{schwing}
The Schwinger-Dyson equations for the partition function of
a rooted branched polymer.}
\end{figure}
$Z$ obeys
\begin{equation}
    Z = \exp(-\mu) \left( 1 + f(Z)\right); ~~~ f(Z) = \sum_{i=2}^{\infty}
w(i)Z^{i-1}.
\end{equation}
Criticality occurs at $\mu = \mu_c$ such that
 $\partial Z / \partial \mu |_{\mu_c} \rightarrow
\infty$ providing that $f$ is convergent in $[0,Z(\mu_c) \equiv Z_c]$;
this criterion is satisfied for the measures that we will consider.
$Z(\mu)$ will then have a leading singularity
of the form $\sim (\mu - \mu_c)^{1/2}$.
     The mean coordination number (minus $1$) equals
\begin{equation}
 \langle q - 1 \rangle = {\frac{\sum_{i>1}^{\infty}
 (i-1)w(i) \partial Z/\partial
   w(i)}{ - \partial Z/ \partial \mu}}.
\end{equation}
By differentiating the Schwinger-Dyson equation with respect to $w(i)$
and $\mu$ and substituting in above,
we find that simply $\langle q-1 \rangle = f'(Z) \exp(-\mu)$
which equals $1 - 1/\langle A \rangle$. $\langle A \rangle$ is
the mean surface area which diverges at $\mu = \mu_c$.
Therefore ${\cal{S}}(\mu)$ (\ref{meansizbp}) is
non-singular for $\mu \ge \mu_c$ and $p < 1$.  Thus criticality
drives the percolation threshold $p_c$ to $1$.
   The partition function for the $q$-state
Potts model on a branched polymer can also be
analyzed in the limit $q \rightarrow 1$
(this was done for the Ising model in (\cite{Ambjmult})).
 From this one can obtain an exact formula for the number of
clusters per unit area, which turns out
to just depend linearly on $p$.

    We close this section by reiterating that the equivalence
of the behavior of percolation clusters on branched polymers
(which we have analyzed) and
on worldsheets in the large $d$ limit relies on unproven assumptions.
It is not perhaps so obvious that percolation on the saddle points
in the space ${\cal{T}}_n$  will behave as on the
saddle points of ${\tilde{\cal{T}}}_n$, which we may not even
have completely characterized.  Thus, these large $d$ results
should be interpreted with appropriate caution.  The preceding arguments
do, in any case, illustrate the behavior of percolation
on a class of triangulations which form tree-like structures.

\section{$p_c = .5$}

    We now turn to an examination of how the geometry of
the lattice constrains the value of the percolation threshold.
Consider the function $K(p) = N_c/N$, the number of clusters
per unit area of the lattice.
Sykes and Essam demonstrated \cite{SykesEss} that on any
planar triangulated multiply-connected lattice,
\begin{equation}
\protect\label{Kreln}
K(p) = K(1-p) + 2p(p - {\frac{1}{2}})(p - 1).
\end{equation}
This relation is exact for finite $N$ for planar lattices; for
lattices with the topology of the torus, it is obeyed up to corrections
of order $1/N$.  The
derivation of (\ref{Kreln}) is based on an application of Euler's
theorem to the two subgraphs consisting of the set of all black (white)
clusters.   The $1/N$ corrections just arise from the
genus term in Euler's equation for subgraphs embedded
on surfaces of higher topology.
The key element of the proof of (\ref{Kreln}) is the following
simple fact:  on any triangulation, each face consists of vertices
which are {\it{all}} connected to each other.

     The identity (\ref{Kreln}) is analogous to the duality
relation for the free energy of the $2-d$ Ising model, though in this
case the geometrical notion of lattice duality is quite different.
We briefly digress to discuss this now.  Consider a lattice
${\hat{\cal{L}}}$
and select a set of faces ${\cal{F}}$.  We then close-pack all faces
${\cal{F}}$; close-packing consists of linking all vertices on a face
together.  This yields a new graph ${\cal{L}}$.  Let
${\cal{L}}^{*}$ then be the lattice
constructed by close-packing all faces in ${\hat{\cal{L}}}$
that are not in ${\cal{F}}$.
Then (\ref{Kreln}) is a special case
of the identity
\begin{equation}
K_{{\cal{L}}}(p) = K_{{\cal{L}}^*}(1-p) + \phi(p)
\end{equation}
where $\phi$ is referred to as the matching polynomial. ${\cal{L}}$ and
${\cal{L}}^{*}$ are matching lattices; any triangulated lattice is
self-matching.

     The analogy with Ising duality has a bit more content, since
if there is a transition at $p=p_c$, $K(p) \sim (p - p_c) ^ {
2 - \alpha}$.  In the mapping of percolation to the q-state Potts model,
$\alpha$ is the exponent that describes the singular behavior of
the free energy (which obeys the Ising duality relation)
and the specific heat.  Thus, we can compute $\alpha$ for percolation
coupled to $c \leq 1$ gravity with the KPZ formula as in (\ref{KPZmsz}),
substituting the weight of the energy operator ($\Delta_{\varepsilon}^o
 = 5/8$) in
the $q \rightarrow 1$ limit of the Potts model for the spin operator
weight of $5/96$.  Then assuming the standard
hyperscaling relation,
one finds that
the dressed weight of the energy operator $\Delta_{\varepsilon}$ equals
$(1-\alpha)/(2-\alpha)$.  For percolation with no matter ($c=0$), the
form of $K(p)$ has been computed by Kazakov
\cite{Kazakov} by analyzing matrix
model saddle point equations; $\alpha = -2$ and $K(p)$ also has
logarithmic contributions to scaling.
 From the KPZ formula, we see that
the exponent $\alpha$ then decreases with $c$, reaching
an irrational value of $-(  2\sqrt{10} + 2)/3  \sim -2.77$
at $c = 1$.

      Thus for $ 0 \leq c \leq 1$, $K(p)$ is always non-analytic at
its critical point, and perforce by (\ref{Kreln}), $p_c = 1/2$.
If the surfaces admit a percolation transition for $c > 1$ with
a value of $\alpha$ that varies continuously with $c$, then $p_c$
must still equal $1/2$.   If we discover that a transition
no longer occurs near $p_c = 1/2$, then we can conclude that
either that there is no percolation transition for $0 < p < 1$ (as
for branched polymers)  or possibly
that a sort of mean-field percolation transition occurs. In the
former case, two types of lattice behavior could occur.
The first possibility would be a collapsed lattice in which virtually
all vertices
are connected to each other; such a lattice would have a finite
probability of containing,
an infinite cluster
for all $p > 0$.  On the other hand, if the average connectivity
were sufficiently low
(as for the finite-genus surfaces constructed from gluing
tetrahedra, discussed in the previous section),
then no infinite clusters would occur for all $p < 1$ \footnote{An example
of an extremely branched triangulated lattice with $p_c = 1$
also is given
in reference \cite{VanBerg}.}.
A `mean-field' transition explicitly refers
a transition in which $\alpha$ is constant and integral
over a range of central charge. For instance, in flat space, the mean
field solution for percolation (above $d=6$) yields $\alpha = -1$;
such a solution would describe trees with sufficiently high mean
connectivity
\footnote{A somewhat pathological example of a triangulated surface
in this universality class with $p_c < 1$ consists
of a lattice of connected tetrahedra with infinite genus
(with $\langle q \rangle > 6$ and $p_c = 3/(\langle q \rangle -3)$)
described by
the construction in the previous section.  The identity (\ref{Kreln})
is no longer true for infinite genus surfaces, however.}.
It is not so clear that one can construct a finite genus surface
surface in the universality class of the Bethe lattice with $p < 1$;
perhaps the mean-field scenario is also not allowed for $1/2 < p < 1$.
In any case, both the mean-field and no-transition scenarios
do seem to be characteristic of percolation on
trees rather than surfaces; they
should signify a qualitative change in the geometry of the worldsheet.

With this background, we now summarize our plan of attack.
We will simulate percolation on $c \leq 1$ worldsheets,
checking that indeed $p_c = 1/2$ and verifying theoretical
predictions for critical behavior.  Assuming $p_c = 1/2$ will allow
us to measure the exponents much more precisely than would usually
be the case. For generically a slight mismeasurement in the
critical temperature will induce a significant error in the finite-size
scaling
determination of critical exponents.  Then to look for a degeneration
of the worldsheet,
we will attempt to determine
if $p_c$ is no longer near $1/2$ for $c > 1$.

\section{The Simulation  \protect\label{SIM}}

      We now provide a discussion of details of our
simulation and analysis.  Our simulations were performed on dynamically
triangulated random surfaces (DTRS), using the standard flip algorithm
\cite{Migdalold} to update the adjacency matrix $C_{ij}$.
The flip algorithm rejected updates that led to
triangulations with loops of length $1$ or $2$ or vertices
with coordination number less than $3$.
We examined DTRS with no embedding coordinates
as well as with Ising spins or Gaussian
fields attached to the
vertices (with interactions along the links).

       Our action for the Ising model was
$S = \beta_c \sum_{(ij)} \sigma_i \sigma_j$.  We chose
a measure independent of $q_i$, as in the matrix model
formulation.
The Ising spins were updated using the Swendsen-Wang cluster
algorithm \cite{SwenWang}.  Each update consisted of
$3N$ attempted flips of randomly chosen links followed by
a Swendsen-Wang update of the entire lattice.
The critical temperature for Ising spins on lattices dual
to our triangulations was computed analytically in
\cite{BurdJur}, so by the standard Ising duality we can determine that
$\beta_c = (1/2)\ln (131/85) \sim .216273$.  This particular
discretization of the Ising model has been studied in
\cite{Jur,ournewpaper}.

For $c=1$ and $2$ we used a standard Metropolis algorithm,
in which the Gaussian fields were updated via random shifts from
a flat distribution.  In the Gaussian case, $3N$ randomly chosen flips
were performed and then $3N$ randomly chosen $X$ coordinates
were updated.  We used the standard Gaussian action with the
conformal measure, which corresponds to the choice $\alpha = 0$
in (\ref{gausspartfn}).  We also simulated the $d=1$ Gaussian
model with the choice $\alpha = -1/2$ and found that the
critical behavior (as determined from the calculation of finite-size
scaling exponents) was identical, within our statistics, to that
measured for $\alpha =0$.

      A percolation measurement began with a coloring
of the lattice with random numbers, produced by a Fibonacci
generator.  All clusters on the lattice were then constructed
(between adjacent sites with random numbers valued less than
$p$) using the method of Hoshen and Kopelman \cite{HoshKopel}.
For smaller lattices and smaller values of the central charge,
percolation clusters were built after each update of the lattice
and matter.
For $c \ge 1$ we found that it was more efficient
to construct clusters after every 5-10 updates of the lattice.
  For $c < 1$, clusters were formed using several
different values of $p$ for each lattice.  For smaller lattices,
the values of these
observables could be extrapolated to nearby values of $p$ using
histogramming techniques \cite{histogrammers}.  Thus, for $c < 1$,
there are strong correlations between the values of observables
at different values of $p$.
Histogramming proved to be less reliable for larger lattices
and for $c \ge 1$.  For $c \ge 1$, our statistics were not quite
so good and we were thus more wary of statistical artifacts
influencing our interpretation of the data.
We therefore performed entirely independent
simulations for each value of $p$, so data points at different $p$
would be uncorrelated.

      Our errors were computed using the jacknife technique.
We first used the jacknife error to
estimate the auto-correlation time, $\tau \sim
1/2 (\sigma_{jack}/\sigma_{naive})^2$; $\sigma_{naive}$ is
just the standard error, assuming no correlations.  Additionally,
we
measured the auto-correlation function, and using standard
methods \cite{Sokal} and appropriate fits, determined
the integrated autocorrelation times (with errors on these
times obtained via jacknife).  Generally, these two techniques
yielded consistent results.  For some of the largest lattices,
the fits to the auto-correlation function were more reliable, since
the jacknife error did not plateau convincingly
as a function of increasing bin-size.  Thus, some of the estimates of
the error
bars for data on the largest lattice ($N=16384$) might be a bit off.

      To obtain the desired precision for our observables, we sampled
a large number of lattices -- between $3 \times 10^4$ and $2.5 \times
10^5$ independent samples, requiring up to $2 \times 10^6$ sweeps
per data point.  We performed the simulations on HP 9000, IBM RS6000
and DEC MIPS workstations; in total we used roughly the equivalent of
$5$ months of time on a HP workstation to gather our data.   Our lattices
were of toroidal topology.  For $c =0$, we used lattices of size $512$
through $16384$; for $c = 1/2$ our lattices ranged in size from $N=2048$
to $16384$.  The correlation times were much longer for $c = 1$ and $2$;
thus the $c = 1$ runs were performed for $N = 1024$ through $8192$
at $p_c =1/2$ and $1024$ and $2048$ for other $p$.  $c = 2$ runs
were only done on lattices of size $1024$ and $2048$ and also at $p=1/2$
using $N=4096$.

    Our basic set of observables consisted of : the number of clusters
per unit volume $K(p)$, the maximum cluster size per configuration
${\cal{M}}(p)$ and the mean cluster size ${\cal{S}}(p)$.

     Our other main task was to determine the critical temperature,
to verify that $p_c = 1/2$ for $c \leq 1$ and then to look for $p_c$
for $c > 1$.  Finding an appropriate technique to do this was not
so straightforward.  To illustrate this difficulty, we first recall
that the width of the transition region is of magnitude
$N^{1/\nu d_H}$.  This width should be, for instance, roughly comparable to
the breadth of the susceptibility peak (the mean cluster size
with the infinite, i.e. largest, cluster excluded) and its
asymptotic shift from $p_c(N =
\infty)$.  For $c \leq 1$, $\nu d_H$ ranges from $4$ to about
$4.7$, indicating that the transition will be extremely broad; for
lattices of several thousand nodes, $\delta p \sim .1$ to $.2$.
In contrast, $\nu d = 2$ for the $2d$ (flat space) Ising model,
which will exhibit a transition that is an order of magnitude
sharper on lattices of comparable size.

    In fact, significant effort has been devoted to developing methods
to
determine percolation thresholds precisely.  Unfortunately, nearly
all of the standard techniques are inapplicable in our case.
They involve determining the fraction of spanning clusters
(see e.g. \cite{Kirkpatrick}) or comparing the outside and inside perimeters of
clusters \cite{Ziff}.  These methods rely on the identification of
a lattice boundary, which we cannot do easily on random lattices.
Other techniques also exploit exact analytic results for small cluster
sizes \cite{Margolina}, which are not available to us in this case.
We therefore ended up adopting a variant of a standard technique
used in spin models,
which relies on a determination of intersections of Binder's cumulants.

Explicitly, we shall consider the quantities
\begin{equation}
\protect\label{binders}
u_{\cal{S}} = 1 - {\frac{ \langle {\cal{S}}^2 \rangle}
{3 ( \langle {\cal{S}} \rangle ) ^2 }}
\end{equation}
and
\begin{equation}
\protect\label{binderm}
u_{\cal{M}} = 1 - {\frac{ \langle {\cal{M}}^2 \rangle}
{3 ( \langle {\cal{M}} \rangle) ^2 }}.
\end{equation}
These quantities should obey (for $c < 1$ at least) a finite-size
scaling form
\begin{equation}
\protect\label{binderfss}
u_{{\cal{S}},{\cal{M}}}(N,p) \asymp f_{{\cal{S}},{\cal{M}}}(z); ~~
  z =  (p - p_c)N^{1/\nu d_H}.
\end{equation}
$f$ is the universal finite-size scaling function and $z$ is the
appropriate scaling variable.  The deviation of these
cumulants from $2/3$ just characterizes
the magnitude of fluctuations of the mean and maximum sizes.
For $p \rightarrow \infty$, $f
\rightarrow 2/3$, since in this phase, the infinite cluster saturates
the system and fluctuations diminish.  Asymptotically,
the cumulants for different values of $N$ will intersect at $p= p_c$
(which is much better than $p \sim p_c \pm N^{1/\nu d_H}$)
where they will be linear and have
slope $f'(0)N^{1/\nu d_H}$.   Therefore, the slopes will not
be particularly large, and they will grow slowly; determining
the points of intersection will require very good statistics.

     We briefly digress to point out how the cumulant $u_
{\cal{S}}$ is related to the standard Binder's cumulant in spin models,
$u = 1 - \langle M^4 \rangle/3 (\langle M^2 \rangle)^2$; $M$ is
the magnetization.  Although
it is true that $\langle {\cal{S}} \rangle = \langle M^2 \rangle$,
in general $u \ne u_{\cal{S}}$.  In the Ising model
(and this should presumably generalize to all $q$-state Potts models),
one can easily derive the relation:
\begin{equation}
 \langle M^4 \rangle = 3 \langle {\cal{S}}^2 \rangle -
{\frac{2}{N}} \langle \ \int n(s)s^4/ \int n(s)s \rangle;
\end{equation}
$n(s)$ is again the cluster distribution function and
the latter term is proportional to the probability that $4$ black points
in the lattice lie together in the same cluster.  The sum of the
terms, which both scale the same way in $N$, is thus
just the integrated $4$-point function of the spin-operator,
with a value that should in principle be calculable analytically.  So
our cumulant just measures one (non-negligible) piece of this $4$-point
function.

\section{Numerical Results}

\subsection{Correlation Times}

     We now turn to the results of our simulations.   First,
we remark that we observed considerable critical slowing down
for ${\cal{M}}(p)$ and ${\cal{S}}(p)$ for all values of $c$.
We found that the values of the auto-correlation exponent $z/d_H:~~
\tau \sim N^{z/d_H}$ were identical, within our statistics, for
${\cal{S}}$ and ${\cal{M}}$.   In the following table we summarize
the values of these exponents.
\begin{table}
\begin{center}
\begin{tabular}{|l|l|l|l|l|} \hline
&  c = 0 & c = 1/2 & c = 1 & c = 2 \\ \hline
$z_{\cal{M}}/d_H$ & .69 (2) & .75 (6) & 1.5 (2) & 1.4 (2) \\ \hline
$z_{\cal{S}}/d_H$ & .70 (2) & .74 (6) & 1.4 (2) & 1.4 (2) \\ \hline
\end{tabular}
\protect\caption[CT_z]{ A summary of the critical slowing down
exponent $z/d_H$ for our runs }
\protect\label{tz}
\end{center}
\end{table}

For $c=0$, the dynamics consists solely of the {\it{local}} link flip
updates.  Since our cluster observables exhibit considerable slowing down,
they reflect large-scale correlations in the geometry of the worldsheet.
In contrast, measurements of the distribution of curvature (e.g.
of $\sum_i | q_i - 6|$ ) do not exhibit measurable critical slowing
down.  $d_H$ has been estimated to be about $2.87$ for $c = 0$
\cite{KawaiNim,Migdal}; this
leads to a value of $z$ of $2.0$.  The value $z/d_H \sim .7$ is not
the largest value that one might anticipate.
For, one would naively guess that the decorrelation of the slowest
critical modes of the surface geometry would be determined by the
spectral dimension, $d_s = 2$ \cite{Migdal}, which characterizes
the probability that a random walker returns to its original site
after time $t$, $P(t) \sim t^{-d_s/2}$.  Thus effectively
a random update would take a time proportional to $N$ (corresponding
to $z/d_H = 1$)
to diffuse across the surface.

 Critical slowing down becomes more severe with increasing $c$.  This
 could be partially due to changes in the fractal properties
of the geometry; presumably the updates of the matter induce additional
slowing down also.   Since the change in $z/d_H$ from $c = 0$ to $c = 1/2$
is rather small, it appears that the Swendsen-Wang algorithm
is quite efficient for our measurements.
Further discussion of the efficiency of spin algorithms on random lattices
will appear in
\cite{csdpaper}.
Our auto-correlation times range from about $\tau_{int} = 1$
(for $c = 0, N= 512$) to $50$ (for $c=2, N= 2048$).  Although
$z/d_H$ is virtually identical for $c=1$ and $c=2$, the correlation times
for $c=2$ are consistently about a factor of $3$ higher than the
corresponding $c=1$ times.  Probably, cluster methods would have been
more efficient in updating the Gaussian fields \cite{lat92c=1}.

We observed one other curious difference between auto-correlation functions
for $c = 0$ and $c > 1$.  For $c = 0$, the fits of the auto-correlation
function (as a function of time $T$)
to the form $\exp (-T/\tau)$
were excellent for $T > 3\tau$ and all values of $N$.
  For $c > 1$, this was no longer true.   The auto-correlation functions
for $c=2$ were quite difficult to fit to; even for times considerably
larger than $\tau$, they exhibited very strong transients.
We shall argue in subsequent sections that for $c > 1$ that we do
not observe good scaling behavior; the behavior of the auto-correlation
functions could perhaps be interpreted as a first hint that this might
be so.

\subsection{Critical Exponents and KPZ}

We now present tables of the effective critical exponents
$(d_f/d_H)_{eff} \equiv \ln({\cal{M}}_{2N}/{\cal{M}}_N)/\ln2$
followed by $(\gamma /\nu d_H)_{eff}
\equiv \ln({\cal{S}}_{2N}/{\cal{S}}_N)/\ln2$ obtained
through finite-size scaling at $p = 1/2$, as defined in section {\ref{SIM}}.

\begin{table}[b]
\begin{center}
\begin{tabular}{|l|l|l|l|l|} \hline
&  c = 0 & c = 1/2 & c = 1 & c = 2 \\ \hline
$N=512$ & .879 (2) &       & .860 (5) &  \\ \hline
$N=1024$ & .880 (3) &&  .861 (5) & .81 (1) \\ \hline
$N=2048$ & .880 (3) & .866 (5) & .859 (7) &.83 (2) \\ \hline
$N=4096$ & .871 (4) & .866 (6) & .841 (8) &  \\ \hline
$N=8192$ & .875 (5) & .852 (8) & &  \\ \hline
theory & .875  & .855 & .772 &  \\ \hline
\end{tabular}
\protect\caption[c1dfdh]{ Measurements of $(d_f/d_H)_{eff}$ compared
with theoretical predictions of the $N \rightarrow \infty$ limit }
\protect\label{dfdh}
\end{center}
\end{table}

\begin{table}[t]
\begin{center}
\begin{tabular}{|l|l|l|l|l|} \hline
&  c = 0 & c = 1/2 & c = 1 & c = 2 \\ \hline
$N=512$ & .774 (2) &  & .744 (6) &  \\ \hline
$N=1024$ & .772 (4) &&  .739 (7) & .68 (2) \\ \hline
$N=2048$ & .768 (5) & .747 (6) & .732 (7) & .67 (3) \\ \hline
$N=4096$ & .753 (5) & .741 (7) & .700 (10) &  \\ \hline
$N=8192$ & .755 (7) & .722 (10) & &  \\ \hline
theory & .75  & .710 & .544 &  \\ \hline
\end{tabular}
\protect\caption[c1d1gam]{ Measurements of $(\gamma/\nu d_H)_{eff}$ compared
with theoretical predictions of the $N \rightarrow \infty$ limit }
\protect\label{d1gam}
\end{center}
\end{table}

The exact values for $\gamma / \nu d_H$, obtained from
(\ref{KPZmsz}) and (\ref{mszfss}) are $3/4, ( 4 - \sqrt{7/2})/3$
and $ 1 - \sqrt{5/24}$ for $c =0 , 1/2$ and $1$ respectively.  We have
again used $d_f/d_H = 1/2(1 + \gamma/\nu d_H)$.
First, we remark that the agreement between the theoretical values
and our measurements for $c < 1$ is surprisingly good, given that
our runs are on medium-sized lattices.  The exponents generally decrease
as a function of increasing lattice size and central charge $c$.
This seems reasonable, since embryonic universes with necks of the
size of the cutoff
become more dominant as $c$ increases; we can see from the theoretical
calculations that this corresponds to a reduction in the critical exponents.
For $N$ not so much larger than the cutoff, the effect
of these bottlenecks should
be not so apparent and then presumably the critical exponents
would be over-estimated.
For $c=0$ on the larger lattices, the exponents have more or less
already leveled off, within our
statistics, to their asymptotic values.

         For $c=1$, if we simply assumed that ${\cal{S}}$ and
${\cal{M}}$
scaled as powers of $N$, then our estimates of the critical exponents
will be very far off.  In fact, the measured values of the
effective $c=1$ exponents agree quite well with the theoretical predictions
for $c = 1/2$.  The logarithmic corrections, though, will shift
the effective exponents far away from their asymptotic values. Using
the relations (\ref{susczn}) and (\ref{partc1}) we see that the first
subleading contribution yields
\begin{equation}
\protect\label{gamnuapprox}
 (\gamma/\nu d_H)_{eff}  = \gamma /\nu d_H + \frac{1 + \gamma/\nu d_H}{\ln N}
 + \cdots.
\end{equation}
In the following plot, we compare our data for $(\gamma / \nu d_H)_{eff}$
at $c=1$
with the theoretical prediction, derived from (\ref{susczn}) and
(\ref{partc1}).  We plot (\ref{gamnuapprox}) with the
dotted curve. The solid curve takes into account all corrections
in (\ref{susczn}) and (\ref{partc1});
\begin{figure}
\epsfxsize=4.5in \epsfbox{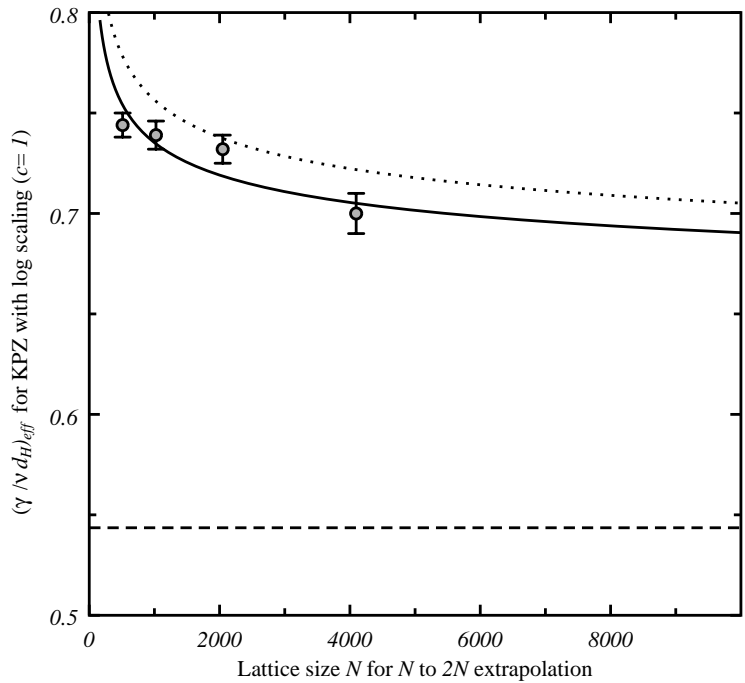}
\protect\caption{\protect\label{gammac1}
A comparison of the theoretical prediction
of the finite size $c=1$ scaling of $\gamma/\nu d_H$ with our data.  The
dotted line includes the leading logarithmic correction and the
solid line takes into account all subleading terms that we calculated. The
horizontal dashed line indicates where these curves asymptote.}
\end{figure}
the smallest correction in the expansion we used to generate this
theoretical prediction was of order $.002$ for $N > 1000$.   At this
order further corrections arising from the logarithmic
renormalization of $\mu$, which we did not compute or conjecture
to be universal,
should appear.
We therefore would guess (without any
rigorous justification) that these additional corrections should
probably shift the curve by less than the
statistical accuracy of our data, which is roughly $.01$.  For $c<1$
we observed small systematic discrepancies from asymptotic scaling,
presumably from power law corrections.
Since power law corrections have also not been taken into
account at $c=1$, we consider
the agreement between our measurements and the both $c=1$ theoretical
curves is thus quite good, certainly comparable to the agreement for
$c < 1$.  This suggests the likelihood that at least the leading $c=1$
logarithmic correction to scaling (\ref{gamnuapprox}) is indeed universal.
We emphasize that we cannot predict the presence of
logarithms in the finite-size scaling ansatz from our data, since it is
not nearly precise enough to distinguish logarithmic behavior
from a small power law.  We can clearly see, though, that if we assume
the logarithmic scaling ansatz given in (\ref{susczn}) and (\ref{partc1}),
that we can extract the proper value of $\gamma/\nu d_H$ from our
data within an accuracy of a few percent.  Also, the agreement
between our data and the KPZ prediction is quite sensitive to the
power with which the leading logs appear in the finite-size scaling
ansatz.

      Naively, it appears that since the effective exponents
for $c=1/2$ and $c=1$ that we measured are quite similar, that
the coupling to matter in both theories also is qualitatively
of the same magnitude.   We see now that this should not be true in
the continuum limit where the exponents are really very different.
One would need to simulate the $c=1$ model on a $400000$
node lattice to see $(\gamma /\nu d_H)_{eff}$ decrease to $.65$, which
would be a somewhat appreciable, though not tremendous difference from
the $c=1/2$ value of $.71$;
$(\gamma /\nu d_H)_{eff} \sim .60$ for $10^{12}$ triangles.
One might think that there is a bit of a conspiracy occurring here:
the $c=1$ KPZ exponents and the logarithmic finite scaling ansatz
somehow mysteriously combine to produce behavior
that strongly resembles $c \sim 1/2$ on lattices
that are numerically accessible to us.  The discussion in
section (\ref{Disc}) should demystify this observation, though.


\subsection{Determining $p_c$}

     We now ascertain if we can successfully estimate $p_c$ from our data.
First, we verify
numerically the identity (\ref{Kreln}),  which constrains $p_c$
to be $0, 1/2$ or $1$ for non-integer $\alpha$.  In figure~
\ref{Kcheck} we have plotted $K(1-p) - K(p)$
(recall that $K$ equals the number of clusters per site)
versus $2p(p-1/2)(1-p)$ using data from $c=0$ simulations with
$N=8192$.  Note that all data points in this plot are heavily correlated;
we used histogramming (from samples taken at $p = .49,.50$ and $.51$)
to produce all of the data points.  Clearly the agreement is excellent,
as expected, since (\ref{Kreln}) should be correct up to corrections of order
$1/N$ for genus $1$.

\begin{figure}
\epsfxsize=4.5in \epsfbox{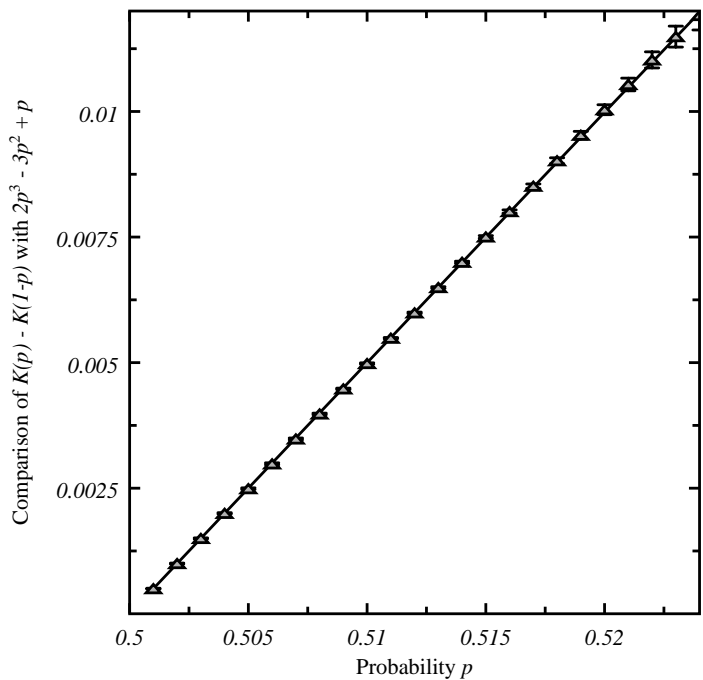}
\protect\caption{\protect\label{Kcheck}A comparison of the
theoretical prediction
relating the the number of clusters per unit area to a polynomial
in $p$.  The data is from $c=0$ simulations on lattices of size
$8192$.}
\end{figure}

       Next, we examine the intersection of the cumulants
$u_{{\cal{M}}}(p)$ and $u_{{\cal{S}}}(p)$ for $c=0$,
presented in figures~\ref{cummaxc0} and \ref{cummszc0} for lattice
sizes $N=1024,2048$ and $4096$.  The curves intersect around
$p_c = .506(4)$ and $.508(4)$ respectively, within a percent or
two of the exact value of $p_c=.5$.  Considering that the large
value of $\nu d_H$ implies that the width
of the transition is of order $.15$,
we conclude that
this technique is fairly successful in locating the critical
point in this case.
\begin{figure}
\epsfxsize=4.5in \epsfbox{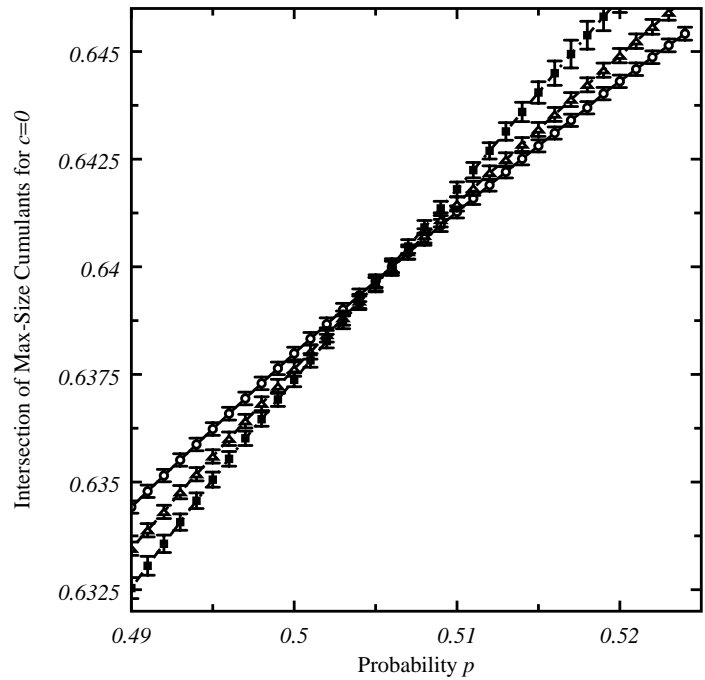}
\protect\caption{\protect\label{cummaxc0}The cumulants $u_{{\cal{M}}}(p)$
for gravity with no matter.  The circles refer to lattice size
$N=1024$,
$N=2048$ points are labeled by triangles, and the
squares correspond to $N=4096$.}
\end{figure}
\begin{figure}
\epsfxsize=4.5in \epsfbox{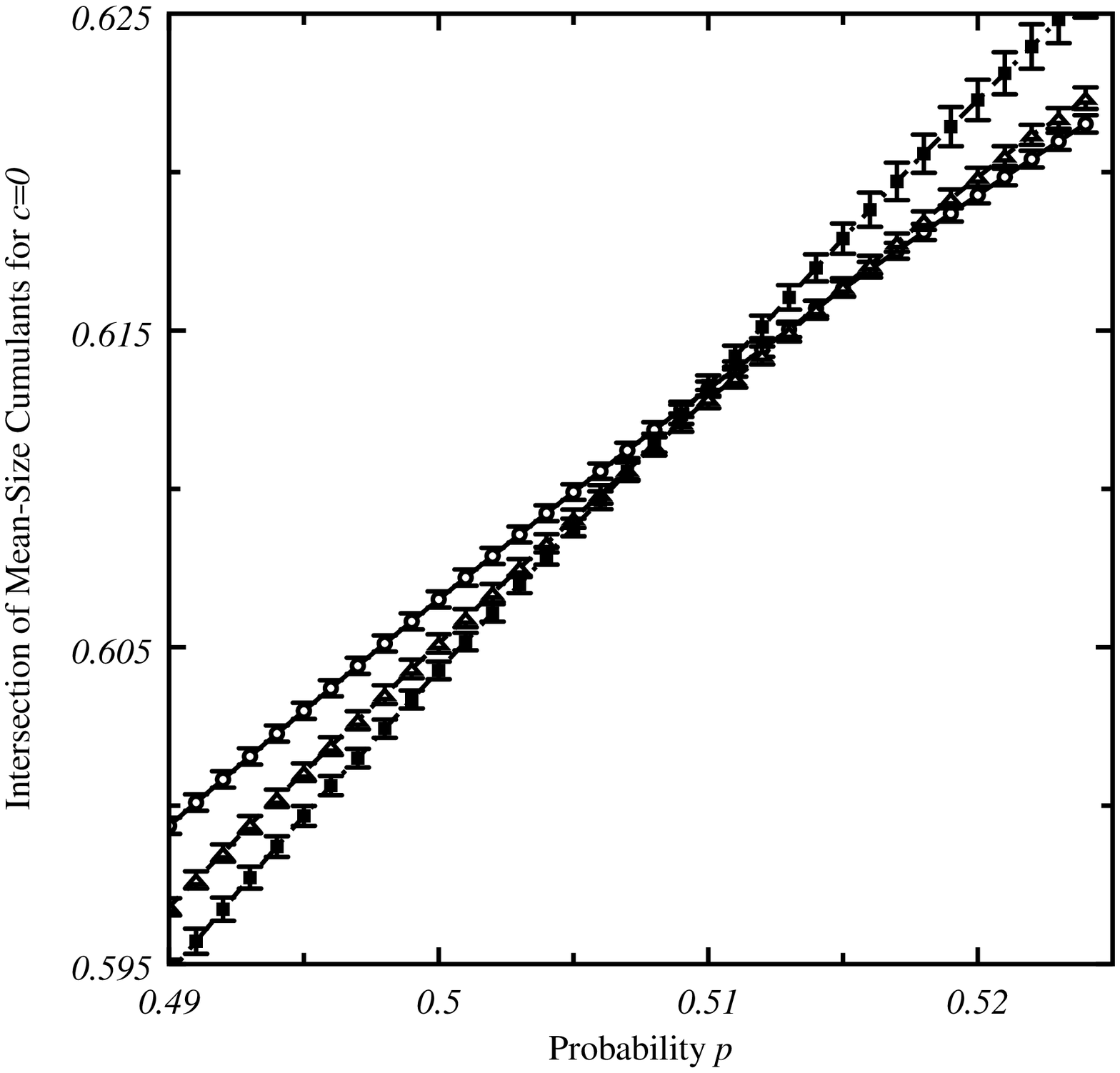}
\protect\caption[cummszc0cap]{\protect\label{cummszc0}The cumulants
$u_{{\cal{S}}}(p)$
for gravity with no matter.  Points are labeled as in the preceding figure.}
\end{figure}
      For $c=1/2$, this technique is at least moderately successful;
the intersection point occurs roughly around $.51$.
We demonstrate this in figure~\ref{cummaxc12},
which shows the curves $u_{{\cal{M}}}$ for $N=2048,4096$ and $8192$.
\begin{figure}
\epsfxsize=4.5in \epsfbox{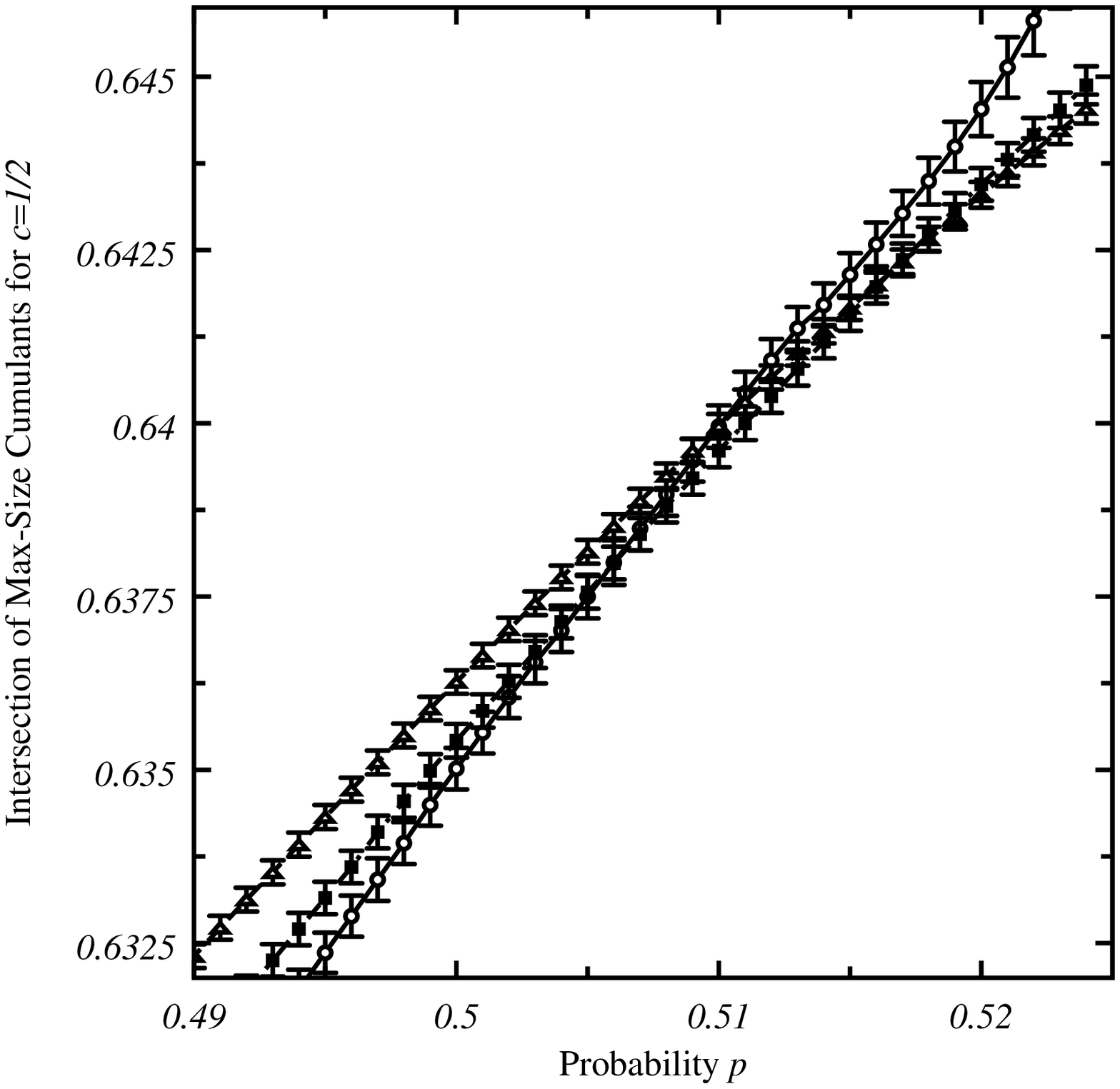}
\protect\caption{\protect\label{cummaxc12}
The cumulants $u_{{\cal{M}}}(p)$
for gravity coupled to $c=1/2$ matter.  Again,
$N=2048$ points are labeled by triangles, $N=4096$ data
is represented by squares; the
circles correspond to
$N=8192$.}
\end{figure}

      At $c=1$, the intersection point of $u_{{\cal{S}}}$,
as shown in figure~\ref{cummszc1},
has moved quite a bit to the
right.  Its position is difficult to determine given our statistics,
but it appears to be  above $p = .52$ and below $.56$.
The data points in this plot are each statistically independent;
we have foregone histogramming because it is no longer so reliable.
Since we know that
$p_c = .5$ from equation (\ref{Kreln}) in this case, we can see
that the method of intersections now has become much less reliable.
\begin{figure}
\epsfxsize=4.5in \epsfbox{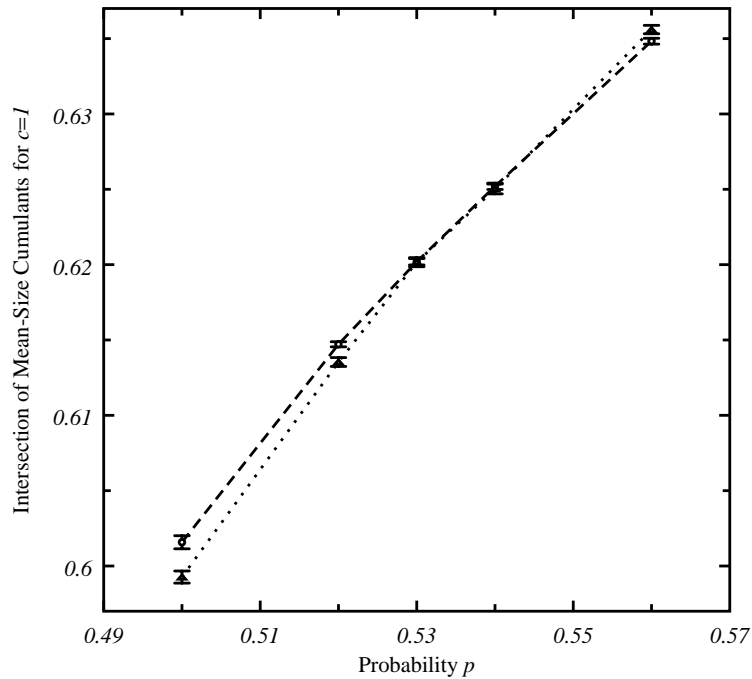}
\protect\caption{\protect\label{cummszc1}The cumulants $u_{{\cal{S}}}(p)$
for gravity coupled to a Gaussian field.
The circles and dashed lines represent $N=1024$
data and
$N=2048$ points are labeled by triangles connected by dotted lines.
Lines are drawn to guide the eye.}
\end{figure}
This failure is expected, though, since the cumulants will only
intersect near $p_c$ when one has reached the scaling regime where
(\ref{binderfss}) is valid.  One would expect that the log corrections
to scaling would modify (\ref{binderfss}) to a relation roughly like
\begin{equation}
\protect\label{binderfsslog}
u_{{\cal{S}},{\cal{M}}}(N,p) \asymp
(1 + {\frac{C}{\ln N}})
 f_{{\cal{S}},{\cal{M}}}(z),
\end{equation}
(the definition of the scaling variable $z$ should also receive logarithmic
corrections).  This modification is anticipated because continuum observables
(such as $f(0)$)
seem to generically receive $1/\ln N$ corrections at $c=1$.  Note that
as $c$ increases and bottlenecks on the worldsheet become more
prominent, the values of the cumulants (at $p=.5$) decrease,
indicating that our observables experience greater fluctuations.
Similarly, increasing $N$ will make the presence of necks more apparent;
this should lead to a decrease in a cumulant, implying that the constant
$C$ in (\ref{binderfsslog}) is positive.  This will then result
in a shift in the intersections to larger values of $p$, as observed.

    For $c=2$, the intersection moves even further to the right.
We illustrate this in figure~\ref{cummszc2}; all points
in this figure are again statistically independent.
\begin{figure}
\epsfxsize=4.5in \epsfbox{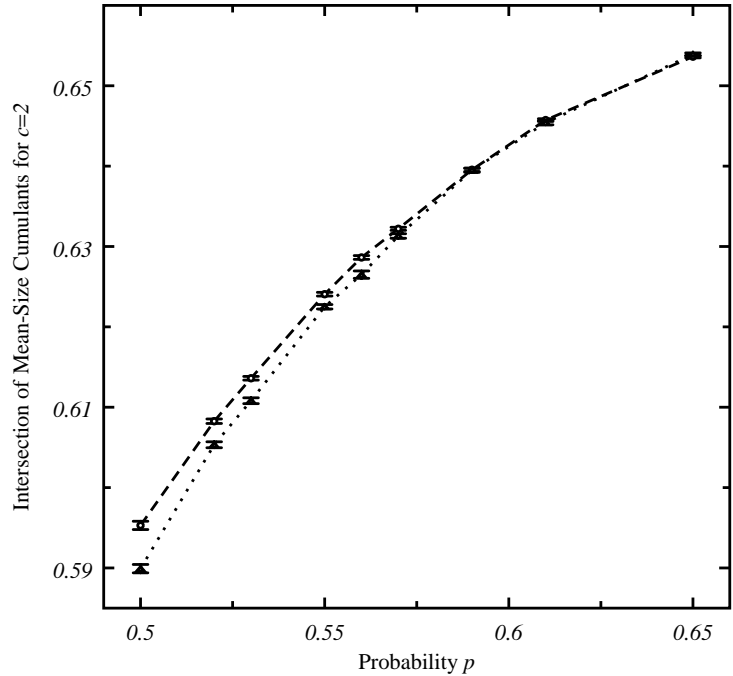}
\protect\caption{\protect\label{cummszc2}The cumulants $u_{{\cal{S}}}(p)$
for gravity coupled to two Gaussian fields.   Points and lines are labeled
as in the preceding figure.}
\end{figure}
In fact,
we cannot determine whether the cumulants actually intersect
or just asymptotically merge, since
they have not even crossed over each other
within our statistics at $p=.65$.   Clearly, though, there is no
intersection between $p = .50$ and $p= .56$.  At first glance,
this appears to be exactly what we were looking for.  Our original
goal was to determine if $p_c$ was appreciable far from $.5$.
If this were true in the continuum, one would infer that either
a transition was absent for $p<1$ or that perhaps the worldsheet
admitted a mean-field percolation transition; either scenario
would be more characteristic of percolation on trees.   One might
hope that the merging of cumulants closer to $.6$ as observed
would just indicate that the percolation correlation
length had finally grown to well beyond the size of our lattices.

     The above scenario, however, is somewhat implausible.  Note that
the effective exponents, e.g. $\gamma/ \nu d_H$ do not change dramatically
at $p= .5$ from $c=1$ and $c=2$.   In general, the $c=2$ and
$c=1$ surfaces do not appear qualitatively so different, so one might
doubt that the percolation correlation length, which is infinite at $p = .5$
and $c=1$ has decreased to roughly the lattice size or below for $c=2$.
We saw that the intersection of the cumulants was shifted because
of finite-size effects at $c=1$.  Presumably finite-size effects
could also
play a large role at $c=2$, so perhaps the large (at least $15 \%$
of $p$!) shift we see in
the cumulants is just a finite lattice artifact.  We shall argue in the
following section that this is indeed probably true and that in fact
any measurements for $c$ somewhat greater than $1$ will
primarily be manifestations of
lattice artifacts.
Let us remark that at least we have observed, via these cumulants,
that the behavior of $c<1$ and $c=2$ is qualitatively different.
The presence of non-analytic behavior of $K(p)$ at $p=p_c$ along
with good scaling behavior on the lattices we consider mandates
that the cumulants intersect near $p=1/2$, as observed for $c < 1$.
This clearly is not the case
for $c=2$.

\section{Discussion \protect\label{Disc}}

      To understand what we are seeing at $c=2$, we return to a discussion
of the scaling corrections
for $c=1$.  A simple physical description of the origin of these
corrections can be found in Klebanov's review \cite{Klebanov}.
We begin by considering the $c=1$ tachyonic operator of momentum $p$:
\begin{equation}
 T(p) \sim \int \sqrt{|{\hat{g}}|}\exp (ipX + (|p|-2)\phi);
\end{equation}
$\phi$ is the Liouville field and ${\hat{g}}$ is the reference metric
($g = {\hat{g}}\exp(-\phi)$).
The Gaussian field obeys \cite{KazMig}
\begin{equation}
\langle X X \rangle \sim (\ln N)^2.
\end{equation}
Thus, the finite lattice size effectively imposes an infrared cutoff
on spacetime momenta of order $1/\ln N$.
The tachyonic energy satisfies
\begin{equation}
      E^2 = p^2 + {\frac{1-c}{12}}.
\end{equation}
Therefore, the minimum ground state energy
accessible at $c=1$ on a lattice of size
$N$ should be roughly
\begin{equation}
      E_{min}^2 \sim (\frac{C}{\ln N})^2,
\end{equation}
for some constant of order unity $C$.
This correction can be quite large; it contributes to the
logarithmic dressing of correlation functions and thus the logarithmic
corrections we have measured that mask the degeneration of the
worldsheet.

      Furthermore, one would anticipate that for finite $N$,
the Gaussian model coupled to gravity should qualitatively resemble
a range of $c < 1$ models.  For the $c < 1$ conformal field theories
can be expressed as continuum limits of lattice models in which
the worldsheet is embedded on Dynkin diagrams \cite{Pasquier}.
$c=0$ corresponds to an embedding into ${\cal{A}}_2$, $c = 1/2$
to ${\cal{A}}_3$ and $c \rightarrow 1$ is identified with the
$n \rightarrow \infty$ limit of ${\cal{A}}_n$.   For fairly large $n$,
the finite worldsheet lattice size imposes an effective cutoff
on these diagrams and screens out the presence of all but a few
Dynkin vertices, producing behavior characteristic
of lower $c$.

      The same mechanism should be in force for $c>1$.  The
extrinsic Hausdorff dimension, $D_H$, may no longer be infinite
but it presumably is large.  If
branched polymers do dominate in
the large $c$ limit, then $D_H(c=1) = \infty$ and
$D_H(c \gg 1) = 4$ \cite{David,Ambdfo} \footnote
{This is the Hausdorff dimension for non-interacting branched
polymers; for low $c$, if the worldsheet is in a branched polymer
phase, interactions should still be relevant and there
would be no compelling reason to believe that $D_H=4$.}.  Numerical
work has indicated that for $c$ of about $2$ or $3$, that $D_H$
is roughly $8-10$ \cite{DavMigold,Jurkold}, though these values
are probably not reliable \footnote{We cannot {\it{a priori}} preclude
the possibility that $D_H$ remains infinite for $c > 1$.}.
For $c > 1$, one would then expect that the minimum ground state
energy accessible on a lattice of size $N$ would be roughly
\begin{equation}
\protect\label{emin}
E_{min}^2 \sim ({\frac{C}{N^{1/D_H(N,c)}}})^2 + {\frac{1-c}{12}},
\end{equation}
where again $C$ is a constant of order unity, perhaps with
a factor of $2 \pi$ thrown in.
For fixed $N$, one would only expect to observe a
gradual increase in the density of
bottlenecks that pinch off large baby universes as $c$ increases
and $E_{min}$ effectively decreases.  A degeneration of the
worldsheet into a polymer-like structure would only become clearly
apparent when $c$ reaches a value such that
(\ref{emin}) is approximately zero.    Numerically, it has
been observed that this degeneration is evident roughly when $c$
is about $10$-$12$
for lattices of order $N = 10^3$.  If we assume that at this point
$D_H = 4$ \cite{Jurkold}, then this determines the
constant $C$ to be about $6 \sim 2 \pi$, an entirely reasonable value.
Of course, these should be only interpreted as back of the envelope
estimates, designed to show that this scenario is qualitatively
consistent with numerical observations.

   At $c=2$, $E_{min}$ reaches $0$ when $N \sim C^{D_H(c=2)}12^{D_H(c=2)/2}$.
Thus, if either $C$ is somewhat greater than $1$ (e.g. $6$ as
in the previous paragraph) or if $D_H(c=2)$ is rather large, one would
need
extraordinarily large lattices to directly
see the degeneration of the worldsheet.  For instance, $C=6$ and
$D_H(c=2)=10$ imply that $N \sim 10^{13}$; $C=1$ and $D_H(c=2)=10$
yield $N \sim 2.5 \times 10^5$.  If $C$ is not larger than $1$ and
$D_H(c=2)$ is smaller (near $4$, for example) then the tachyonic degeneration
should be observable on the lattices we consider and finite-size
effects should not be so dominant.  These values, however,
would then not be consistent with previous much larger measurements
of $D_H(c=2)$.
Again, we do not advocate taking particular numbers too seriously.
It is apparent, though, that the true continuum behavior at $c=2$
might only become evident at scales orders of magnitude larger than
those that are amenable to simulation.

    One would then expect also that the percolation correlation length
at $p = .5$ (where a transition would occur for surfaces) should
be comparable to the length scale at which the surface degenerates
into a branched polymer.  If we accept the possibility that this
length scale is very large, then
for $c=2$, we should not be able to detect
the presence of this correlation length
on surfaces of a few thousand nodes.  From the relation (\ref{emin})
it follows that we should observe significant deviations from scaling,
since the shift in the effective ground state mass is detectable
as a function of $N$ (as at $c=1$).  In fact,  $1/\ln(10^3) < 1/(10^{3/D})$
for $D > 4$, so for the range of $N$ that we simulate, the shifts
in finite-size corrections could be larger for $c = 2 $ than for $c=1$.
Therefore, the large shift in the cumulant intersections at $c=2$
is most likely just a manifestation of a lack of good scaling behavior.
All other possibilities do not fit very well with at least some
portion of the prevailing numerical evidence.  If at $c=2$ the worldsheet
had degenerated into a tree-like structure, and our measurements
were indeed satisfactorily measuring asymptotic scaling behavior,
then some evidence of the tachyonic degeneration should have been
easily apparent in previous simulations.  If on the other hand,
no such surface degeneration had occurred and our measurements
were within the scaling regime, then we would have not anticipated
a large shift in the cumulant intersection from $p = .5$.

     This leads us to a rather disappointing conclusion.  The above
arguments suggest the likelihood
that generically numerical simulations for $c$
somewhat greater than one will just measure lattice artifacts.  This
has also been borne out by previous attempts to measure $\gamma_s$,
which indicated that corrections to scaling were predominant
for $c \geq 1$ \cite{Ambgs}.  On lattices
of accessible size, the onset of the tachyon could be hidden by finite-size
effects, and we would fail even to capture the qualitative nature of
these models in the continuum limit. Scaling exponents
and critical properties of the geometry (the Hausdorff dimension, e.g.)
measured in these simulations would not describe
continuum scaling behavior.  For very large $c$, simulations
may reflect the character of the tachyonic instability, however.
At $c=1$, the results of simulations can only be properly understood
if one knows the form of the corrections to scaling, which can be computed
because the theory is solvable.  For $c < 1$,  we have found that
numerical simulations do reproduce, with reasonable precision, critical
exponents that characterize the behavior of percolation on
strings.

\section{Acknowledgments}
       The numerical simulations done in this work employed
DTRS software originally written by Enzo Marinari.  Leping Han and
Marco Falcioni also later contributed to the development of this code.
Much of the data analysis was performed using programs supplied by
Paul Coddington and Enzo Marinari.  This project was done using
NPAC (Northeast Parallel Architecture Center) and CASE center
computing facilities.   During the course of this work,
I enjoyed enlightening conversations and correspondence with
Mark Bowick, Paul Coddington, Marco Falcioni, Gerard Jungman,
Volodya Kazakov and Enzo Marinari.  I am grateful to all of the above
people for their assistance and encouragement.  This work was sponsored
by Department of Energy
Grant DOE DE-FG02-85ER40231 and research funds from Syracuse University.

\end{document}